# Stability Analysis of A Feedback-linearization-based Controller with Saturation: A Tilt Vehicle with The Penguin-inspired Gait Plan

Zhe Shen, Yudong Ma, Takeshi Tsuchiya
*Department of Aeronautics and Astronautics, The University of Tokyo, Japan*

**Abstract**

Saturations in control signal can challenge the stability proof of a feedback-linearization-based controller, even leading the system unstable [1]. Thus, several approaches are established to avoid reaching the saturation bound [2,3]. Meanwhile, to help design the controller for a quad-tilt-rotor, [1] modeled a tilt vehicle with implementing the feedback-linearization-based controllers. In this article, we provide a gait plan for this tilt vehicle and control it utilizing the feedback linearization. Since saturations exist in the control signals, we study the stability based on Lyapunov theory.

**Keywords**
Feedback linearization, Gait plan, Lyapunov, Stability

**1. Introduction**

[1] put forward a tilt vehicle to aid the controller designation of the tilt rotor [4]. As reported, the gait plan is employed to avert the 'State Drift' phenomenon. Gait plan is originally adopted in four-legged robots [5-7], which was inspired by the four-legged animals. Parallel to the concept, we perform the gait plan for the tilt vehicle based on the moving pattern of penguins, swinging to left and right periodically.

The procedure following gait plan is feedback linearization; this technique converts the nonlinear system to a linear system, providing the possibility of applying the linear controllers. This method is widely accepted in the studies [8-11] of the quadrotor control.

The generally adopted protocol in feedback linearization is averting the saturations. This is because that the system is no longer linearized when the states saturate. Adopting the same control rule for the system with saturation challenges the stability proof, leading to unstable in some systems [1]. Several researches focus on avoiding activating the saturation constraints [12-16]. While the studies focusing on the stability proof of the dynamic inversion with saturation are rare.

In this research, we plan two gaits for the tilt vehicle. In the first gait, the feedback linearization with non-saturation states is assured. In the other gait, the saturation in states is inevitable, defying the previous stability proof (stability proof of the linear controller). To guarantee the stability of our controller, we provide the stability proof of the dynamic inversion with saturation for our system with the straight-line reference.

The Lyapunov criteria are commonly used in proving the asymptotically stable in UAV controls. Several well-known controllers are designed based on it: geometric control [17-20], backstepping control [21-23], sliding mode control [24-26], etc. Lyapunov candidates are found negative-definite in these studies, indicating the asymptotically stable [27,28]. However, seeking a Lyapunov candidate with such a requirement is hard or even impossible. Instead, we utilize the Lyapunov theorem to prove the definition of stability below.

$\exists \epsilon, \ \exists \delta:$

$if \ |x(0)| < \delta$

$then \ |x(t,x(0))| < \epsilon, \ \forall t \geqslant 0$

This stability tells us: if the system is stable and the initial condition is within the admissible range, then the output is bounded.

The rest of the paper is organized as follows: Section 2 introduces the dynamics of the tilt vehicle. The penguin-gait-based gait plan is presented in Section 3. Section 4 designs the controller with the first gait, which avoids the saturation in states. The stability proof of this controller is also provided in this section. In Section 5, we adopt the same controller developed in Section 4. However, the gait in Section 5 leads to the saturation in states. The stability proof of this case is given in Section 6. The simulation result for this case is presented in Section 7. Section 8 makes conclusions and discussions.

**2. Dynamics of the Tilt Vehicle**

The tilt vehicle [1] to control is sketched in Figure 1—2. Two thrusts are provided by the propellers on the top disc. The top disc can tilt, giving the possibility of altering the direction of the thrusts. The friction between the bottom disc and the ground is neglected.

[1] analyzed its dynamics, which can be summarized in Equation (1).

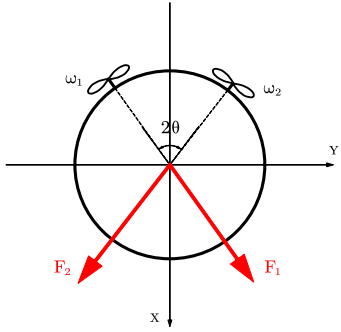

Fig.1. Tilt vehicle

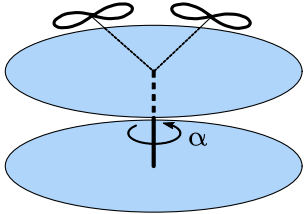

Fig.2. Tilt structure

$$\begin{bmatrix} \ddot{x} \\ \ddot{y} \end{bmatrix} = \frac{1}{m} \cdot J_\Lambda \cdot J_\theta \cdot \begin{bmatrix} \omega_1^2 \\ \omega_2^2 \end{bmatrix} \quad (1)$$

where $m$ is the mass of the entire vehicle. $J_\Lambda$ and $J_\theta$ are defined in Equation (2)—(3).

$$J_\Lambda = \begin{bmatrix} \cos(\Lambda) & -\sin(\Lambda) \\ \sin(\Lambda) & \cos(\Lambda) \end{bmatrix} \quad (2)$$

$$J_\theta = \begin{bmatrix} \cos(\theta) & 0 \\ 0 & \sin(\theta) \end{bmatrix} \cdot \begin{bmatrix} K_{F_1} & K_{F_2} \\ K_{F_1} & -K_{F_2} \end{bmatrix} \quad (3)$$

where $\Lambda$ is yaw angle. $\theta$ is the half of the angle between two propellers (see Fig. 1). In our model, we design theta in Equation (4). $K_{F_i}$ is the thrust coefficient (Equation (5)).

$$\theta = \frac{\pi}{6} \quad (4)$$

$$K_{F_1} = K_{F_2} = K = 0.001 \quad (5)$$

A positive $\Lambda$ represents an anti-clockwise yaw (Figure 3). A negative $\Lambda$ represents a clockwise yaw (Figure 4).

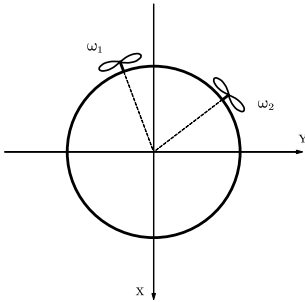 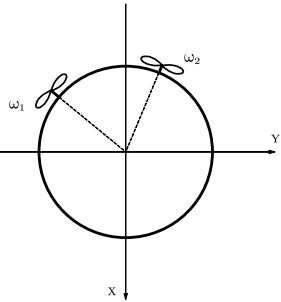

Fig. 3. Tilt to right     Fig. 4. Tilt to left

## 3. Reference and Penguin Gait Based Gait Plan

In this study, we aim to solve the straight-line tracking problem. The reference is a straight line with a constant acceleration. It is defined in Equation (6).

$$\begin{cases} x_r = \frac{1}{2} \cdot t^2 \\ y_r = 0 \end{cases} \quad (6)$$

To mimic the walking pattern of a penguin, we create two gaits planned in Figure 5 – Figure 6.

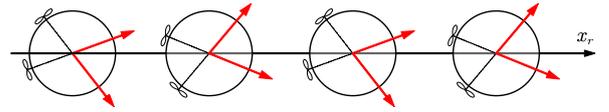

Fig. 5. Penguin-inspired gait with small swing.

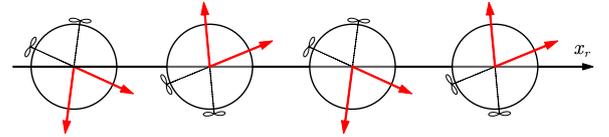

Fig. 6. Penguin-inspired gait with large swing.

Both Figure 5 and Figure 6 plot the swinging-like gait; the yaw of the tilt vehicle changes periodically. The period of it is 2 seconds.

The yaw in Figure 5 changes relatively small, bringing the possibility of maintaining 0 acceleration along $Y-axis$. When yaw is large (Figure 6), it is impossible to receive 0 dynamic state error while maintaining 0 acceleration along $Y-axis$.

The $\Lambda-t$ (yaw-time) history for each gait is in Figure 7 ($|\Lambda| = \frac{\pi}{8}$) and Figure 8 ($|\Lambda| = \frac{\pi}{3}$), respectively.

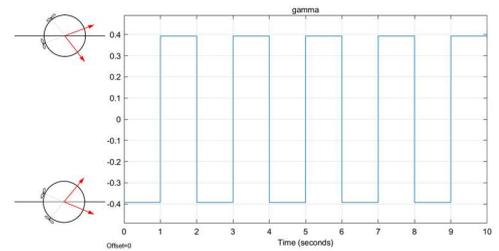

Fig. 7. Small gait with $|\Lambda| = \frac{\pi}{8}$

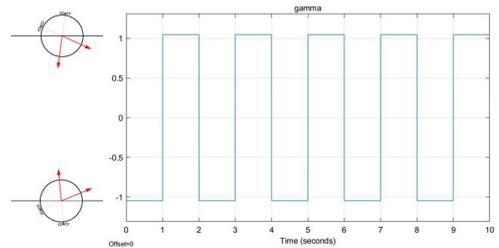

Fig. 8. Large gait with $|\Lambda| = \frac{\pi}{3}$

The small-gait based controller is discussed in Section 4. While the large-gait based controller is discussed in Section 5 – Section 7.

## 4. Gait without Saturation and Control

In this section, the gait planned in Figure 7 ($|\Lambda| = \frac{\pi}{8}$) is employed. The feedback linearization is applied before applying PD controllers.

### 4.1. Feedback Linearization

Notice that $J_\Lambda$ and $J_\theta$ in (1) are invertible. We can separate the inputs from the dynamics in (7).

$$\begin{bmatrix} \omega_1^2 \\ \omega_2^2 \end{bmatrix} = J_\theta^{-1} \cdot J_\Lambda^{-1} \cdot m \cdot \begin{bmatrix} \ddot{x} \\ \ddot{y} \end{bmatrix} \tag{7}$$

### 4.2. PD Controllers

Substituting the desired accelerations into (7) yields Equation (8).

$$\begin{bmatrix} \omega_1^2 \\ \omega_2^2 \end{bmatrix} = J_\theta^{-1} \cdot J_\Lambda^{-1} \cdot m \cdot \begin{bmatrix} \ddot{x}_d \\ \ddot{y}_d \end{bmatrix} \tag{8}$$

We design the PD controllers based on Equation (8). The controllers are in Equation (9) – Equation (10).

$$\ddot{x}_d = \ddot{x}_r + K_{X_1} \cdot (\dot{x}_r - \dot{x}) + K_{X_2} \cdot (x_r - x) \tag{9}$$

$$\ddot{y}_d = \ddot{y}_r + K_{Y_1} \cdot (\dot{y}_r - \dot{y}) + K_{Y_2} \cdot (y_r - y) \tag{10}$$

where $K_{X_1} = 12$, $K_{X_2} = 6$, $K_{Y_1} = 9$, $K_{Y_2} = 18$.

Substitute Equation (9), (10) into Equation (8) yields Equation (11).

$$\begin{bmatrix} \omega_1^2 \\ \omega_2^2 \end{bmatrix} = J_\theta^{-1} \cdot J_\Lambda^{-1} \cdot m$$
$$\cdot \begin{bmatrix} \ddot{x}_r + K_{X_1} \cdot (\dot{x}_r - \dot{x}) + K_{X_2} \cdot (x_r - x) \\ \ddot{y}_r + K_{Y_1} \cdot (\dot{y}_r - \dot{y}) + K_{Y_2} \cdot (y_r - y) \end{bmatrix} \tag{11}$$

However, the right side of Equation (11) is not assured to be nonnegative. While the left side of Equation (11) is the square of the angular velocity, which should be nonnegative.

Thus, rather than direct the right side of Equation (11) to the angular velocities squared, we require zero lower bounds (constraints) at zero. The specific control rule is in Equation (12), (13).

**Practical Control Law (with Saturation)**

$$\begin{bmatrix} squared\omega_1 \\ squared\omega_2 \end{bmatrix} = J_\theta^{-1} \cdot J_\Lambda^{-1} \cdot m$$
$$\cdot \begin{bmatrix} \ddot{x}_r + K_{X_1} \cdot (\dot{x}_r - \dot{x}) + K_{X_2} \cdot (x_r - x) \\ \ddot{y}_r + K_{Y_1} \cdot (\dot{y}_r - \dot{y}) + K_{Y_2} \cdot (y_r - y) \end{bmatrix} \tag{12}$$

$$\begin{bmatrix} \omega_1^2 \\ \omega_2^2 \end{bmatrix} = \begin{bmatrix} \max(squared\omega_1, 0) \\ \max(squared\omega_2, 0) \end{bmatrix} \tag{13}$$

**Remark 1**

When the right side of Equation (12) is positive ($squared\omega_1 \geq 0$, $squared\omega_2 \geq 0$), Equation (12), (13) are equivalent to Equation (11).

The rest of the article applies the controller specified in Equation (12), (13).

### 4.3. Stability Proof (no saturation case)

This section presents the proof of the following fact:

**Proposition 1**

If the control law does not reach the saturation bound for the entire time, that is

$$\begin{cases} squared\omega_1(t) \geq 0 \\ squared\omega_2(t) \geq 0 \end{cases} \quad (t \geq 0) \tag{14}$$

Then, $e_x$, $\dot{e}_x$, $e_y$, $\dot{e}_y$ are bounded applying the controller in Equation (12), (13).

**Proof**

Since the control law does not touch the saturation bound, the controller can be reduced to Equation (11). Substituting Equation (11) into Equation (1) yields Equation (15).

$$\begin{bmatrix} \ddot{x} \\ \ddot{y} \end{bmatrix} = \frac{1}{m} \cdot J_\Lambda \cdot J_\theta \cdot J_\theta^{-1} \cdot J_\Lambda^{-1} \cdot m$$
$$\cdot \begin{bmatrix} \ddot{x}_r + K_{X_1} \cdot (\dot{x}_r - \dot{x}) + K_{X_2} \cdot (x_r - x) \\ \ddot{y}_r + K_{Y_1} \cdot (\dot{y}_r - \dot{y}) + K_{Y_2} \cdot (y_r - y) \end{bmatrix} \tag{15}$$

Rearranging Equation (15) yields Equation (16).

$$\begin{cases} \ddot{e}_x + K_{X_1} \cdot \dot{e}_x + K_{X_2} \cdot e_x = 0 \\ \ddot{e}_y + K_{Y_1} \cdot \dot{e}_y + K_{Y_2} \cdot e_y = 0 \end{cases} \tag{16}$$

where $e_x = x_r - x$, $e_y = y_r - y$.

Define the following Lyapunov candidate:

$$\mathscr{L}_y = \frac{1}{2} \cdot \dot{e}_y^2 + \frac{1}{2} \cdot K_{Y_2} \cdot e_y^2 \tag{17}$$

Differentiating Equation (17) and substituting Equation (16) yield Equation (18).

$$\dot{\mathscr{L}}_y = -K_{Y_1} \cdot \dot{e}_y^2 \leq 0 \tag{18}$$

Similarly, define the following Lyapunov candidate:

$$\mathcal{L}_x = \frac{1}{2} \cdot \dot{e}_x{}^2 + \frac{1}{2} \cdot K_{X_2} \cdot e_x{}^2 \quad (19)$$

Calculate the derivative of (19). We receive (20).

$$\dot{\mathcal{L}}_x = -K_{X_1} \cdot \dot{e}_x{}^2 \leq 0 \quad (20)$$

Thus, $e_x$, $\dot{e}_x$, $e_y$, $\dot{e}_y$ are bounded.

**Remark 2**

For a LTI system, BIBO stable is equivalent to asymptotic stable. While this result is unnecessary.

Moreover, there are several alternative ways to prove asymptotically stable for this LTI system. We use Lyapunov methods since the Lyapunov candidate in Equation (17) will facilitate to deal with the stability problem in Section 6.

4.4. Simulation Result

We simulate the process by Simulink. The Simulink diagram can be found in Figure 9, with several major parts marked.

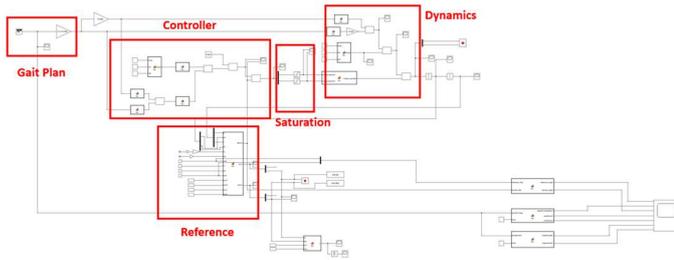

Fig. 9. The Simulink block diagram.

The dynamic state error ($e_x$, $e_y$) is near zero (less than $10^{-10}$) all the time.

The inputs ($\omega_1{}^2$ and $\omega_2{}^2$) are plotted in Figure 10. The inputs change periodically according to the periodical change of the planned gait (yaw signal). Notice that both inputs are positive without touching the saturation bounds. The stability proof holds for this system.

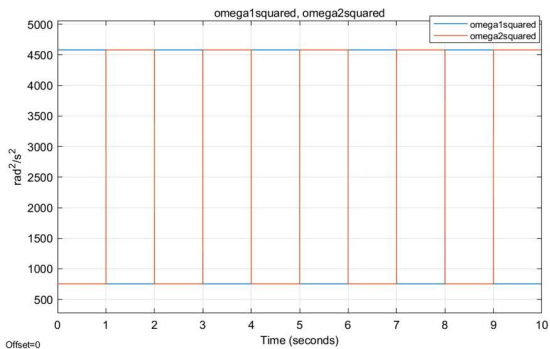

Fig. 10. The inputs.

**5. Gait with Saturation and Control**

In this section, the gait planned in Figure 8 ($|\Lambda| = \frac{\pi}{3}$) is employed. As explained, this gait introducesthe dynamic state error, even if the time is large enough.

We employ the same controller, Practical Control Law, illustrated in Equation (12), (13) to stabilize the tilt vehicle in this gait.

**Proposition 2**

If the gait planned in Figure 8 ($|\Lambda| = \frac{\pi}{3}$) is employed,

Then, $e_x$, $\dot{e}_x$, $e_y$, $\dot{e}_y$ are bounded applying the controller in Equation (12), (13).

Since the saturation bounds are touched in this gait, Proposition 1 is no longer applicable to this case. In other words, the stability is challenged when saturation appears.

In Section 6, we provide the proof of Proposition 2.

**6. Proof of Proposition 2**

**Lemma 1 (Switch Matrix Based Control Rule)**

The practical control rule described in (12), (13) is equivalent to Equation (21).

$$\begin{bmatrix} \omega_1{}^2 \\ \omega_2{}^2 \end{bmatrix} = S_{pq} \cdot J_\theta{}^{-1} \cdot J_\Lambda{}^{-1} \cdot m$$
$$\cdot \begin{bmatrix} \ddot{x}_r + K_{X_1} \cdot (\dot{x}_r - \dot{x}) + K_{X_2} \cdot (x_r - x) \\ \ddot{y}_r + K_{Y_1} \cdot (\dot{y}_r - \dot{y}) + K_{Y_2} \cdot (y_r - y) \end{bmatrix} \quad (21)$$

where $S_{pq}$ is called 'Switch Matrix', which is specified in Equation (22).

$$S_{pq} = \begin{bmatrix} p & 0 \\ 0 & q \end{bmatrix} \quad (p = 0, 1 \quad q = 0, 1) \quad (22)$$

where $p$ and $q$ are determined in (23), (24).

$$p = \begin{cases} 0, & \text{when } squared\omega_1 \leq 0 \\ 1, & \text{when } squared\omega_1 > 0 \end{cases} \quad (23)$$

$$q = \begin{cases} 0, & \text{when } squared\omega_2 \leq 0 \\ 1, & \text{when } squared\omega_2 > 0 \end{cases} \quad (24)$$

where $squared\omega_i$ ($i = 1, 2$) are specified in (12).

**Lemma 2 (Determine Switch Matrix)**

Define $(m, n)$ in Equation (25).

$$\begin{bmatrix} m \\ n \end{bmatrix} = J_\Lambda{}^{-1} \cdot \begin{bmatrix} \ddot{x}_d \\ \ddot{y}_d \end{bmatrix} \quad (25)$$

where $J_\Lambda$ is determined in Equation (2). $J_\Lambda$ is the rotation transformation of $\Lambda$. In this section, we analyze

$|\Lambda| = \dfrac{\pi}{3}$ (large gait). $(\ddot{x}_d, \ddot{y}_d)$ is the desired acceleration, which is defined in Equation (26).

$$\begin{bmatrix} \ddot{x}_d \\ \ddot{y}_d \end{bmatrix} = \begin{bmatrix} \ddot{x}_r + K_{X_1} \cdot (\dot{x}_r - \dot{x}) + K_{X_2} \cdot (x_r - x) \\ \ddot{y}_r + K_{Y_1} \cdot (\dot{y}_r - \dot{y}) + K_{Y_2} \cdot (y_r - y) \end{bmatrix} \quad (26)$$

Then, $S_{pq}$ is determined based on $(m, n)$. The relationship is in Figure 11.

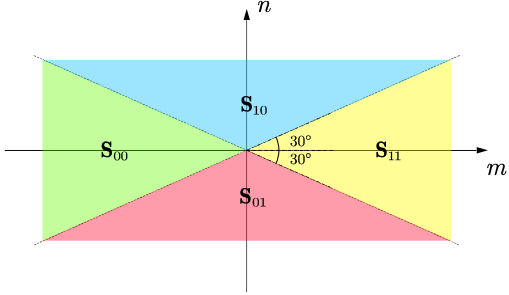

Fig. 11. Relationship between $(m, n)$ and $S_{pq}$.

**Proof**

We will determine $S_{pq}$ based on Equation (12), (23) and (24).

Substituting Equation (3) into Equation (12) yields Equation (27).

$$\begin{bmatrix} squared\omega_1 \\ squared\omega_2 \end{bmatrix} = \dfrac{m}{K} \cdot \begin{bmatrix} \dfrac{1}{2} & \dfrac{1}{2} \\ \dfrac{1}{2} & -\dfrac{1}{2} \end{bmatrix} \cdot \begin{bmatrix} \dfrac{1}{\cos(\theta)} & 0 \\ 0 & \dfrac{1}{\sin(\theta)} \end{bmatrix}$$
$$\cdot J_\Lambda^{-1} \cdot \begin{bmatrix} \ddot{x}_d \\ \ddot{y}_d \end{bmatrix} \quad (27)$$

Define $(u, v)$ in Equation (28).

$$\begin{bmatrix} u \\ v \end{bmatrix} = \begin{bmatrix} \dfrac{1}{\cos(\theta)} & 0 \\ 0 & \dfrac{1}{\sin(\theta)} \end{bmatrix} \cdot J_\Lambda^{-1} \cdot \begin{bmatrix} \ddot{x}_d \\ \ddot{y}_d \end{bmatrix} \quad (28)$$

The relationship between $(u, v)$ and $S_{pq}$ can be identified based on Equation (23), (24), (27) and (28). This result is plotted in Figure 12.

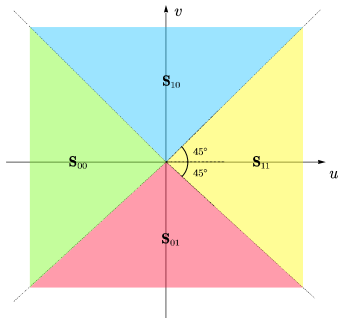

Fig. 12. Relationship between $(u, v)$ and $S_{pq}$.

Notice the relationship between $(u, v)$ and $(m, n)$ in Equation (29).

$$\begin{bmatrix} u \\ v \end{bmatrix} = \begin{bmatrix} \dfrac{1}{\cos(\theta)} & 0 \\ 0 & \dfrac{1}{\sin(\theta)} \end{bmatrix} \cdot \begin{bmatrix} m \\ n \end{bmatrix} \quad (29)$$

Equation (29) is the extension transformation of the coordinates. Its result is exactly Figure 11.

**Lemma 3 (Sufficient Condition for $e_x \equiv 0$)**

$$\text{If} \begin{cases} e_x(0) = 0, \dot{e}_x(0) = 0 \\ \text{when } \Lambda(t) = -\dfrac{\pi}{3}, S_{pq} = \{S_{11}, S_{10}\} \\ \text{when } \Lambda(t) = \dfrac{\pi}{3}, S_{pq} = \{S_{11}, S_{01}\} \end{cases},$$

$$t \in \left[0, n \cdot \dfrac{T}{2}\right] \; (n \in \mathbf{Z}^+, T \text{ is the period } (2s))$$

$$\text{then} \quad e_x \equiv 0, \quad t \in \left[0, n \cdot \dfrac{T}{2}\right]$$

**Proof**

1° Calculate the dynamic state error in $S_{11}$

Substituting $S_{11}$ and Equation (21) into Equation (1) yields Equation (30).

$$\begin{cases} \ddot{e}_x = -K_{X_1} \cdot \dot{e}_x - K_{X_2} \cdot e_x \\ \ddot{e}_y = -K_{Y_1} \cdot \dot{e}_y - K_{Y_2} \cdot e_y \end{cases} \quad (30)$$

2° Calculate the dynamic state error in $S_{10}$

Substituting $S_{10}$, Equation (21), and $\Lambda(t) = -\dfrac{\pi}{3}$ into Equation (1) yields Equation (31).

$$\begin{cases} \ddot{e}_x = -K_{X_1} \cdot \dot{e}_x - K_{X_2} \cdot e_x \\ \ddot{e}_y = \dfrac{1}{\sqrt{3}} \cdot (K_{X_2} \cdot e_x + K_{X_1} \cdot \dot{e}_x + 1) \end{cases} \quad (31)$$

3° Calculate the dynamic state error in $S_{01}$

Substituting $S_{01}$, Equation (21), and $\Lambda(t) = \dfrac{\pi}{3}$ into Equation (1) yields Equation (32).

$$\begin{cases} \ddot{e}_x = -K_{X_1} \cdot \dot{e}_x - K_{X_2} \cdot e_x \\ \ddot{e}_y = -\dfrac{1}{\sqrt{3}} \cdot (K_{X_2} \cdot e_x + K_{X_1} \cdot \dot{e}_x + 1) \end{cases} \quad (32)$$

Based on the dynamics of $e_x$ in Equation (30) – (32), $e_x \equiv 0$ given that the initial condition is zero.

**Lemma 4 (Iteration in $S_{pq}$)**

If $e_x \equiv 0$, $t \in \left[0, n \cdot \dfrac{T}{2}\right]$,

$(n \in \mathbf{Z}^+, T$ is the period $(2s))$

then $\begin{cases} \text{when } \Lambda(t) = -\dfrac{\pi}{3}, S_{pq} = \{S_{11}, S_{10}\} \\ \text{when } \Lambda(t) = \dfrac{\pi}{3}, S_{pq} = \{S_{11}, S_{01}\} \end{cases}$,

$t \in \left[n \cdot \dfrac{T}{2}, (n+1) \cdot \dfrac{T}{2}\right]$

**Proof**

Substituting $e_x \equiv 0$ into Equation (26) yields (33).
$$\ddot{x}_d \equiv 1 \tag{33}$$

Notice that $J_\Lambda^{-1}$ in Equation (25) is a rotational matrix. It rotates $[\ddot{x}_d \; \ddot{y}_d]'$ by $-\Lambda$ to $[m \; n]'$. Equation (33) restricts the position of $[\ddot{x}_d \; \ddot{y}_d]'$, before the rotation transformation. We present the possible $[\ddot{x}_d \; \ddot{y}_d]'$ in Figure 13 (red dash line).

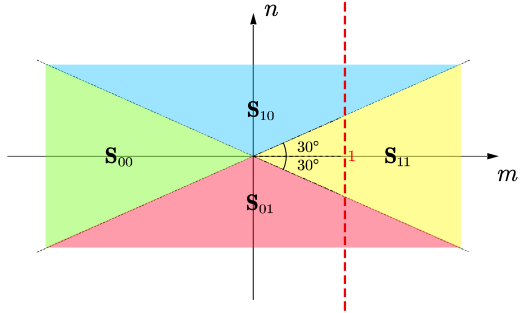

Fig. 13. The possible positions of $[\ddot{x}_d \; \ddot{y}_d]'$ are on the red dash line.

1° Find $S_{pq}$ when $\Lambda(t) = -\dfrac{\pi}{3}$

In this case, we need to rotate $[\ddot{x}_d \; \ddot{y}_d]'$ by $\dfrac{\pi}{3}$ $(-\Lambda(t))$ to receive $[m \; n]'$.

It can be found that $[m \; n]'$ can only lie in $S_{10}$ or $S_{11}$.

2° Find $S_{pq}$ when $\Lambda(t) = \dfrac{\pi}{3}$

In this case, we need to rotate $[\ddot{x}_d \; \ddot{y}_d]'$ by $-\dfrac{\pi}{3}$ $(-\Lambda(t))$ to receive $[m \; n]'$.

It can be found that $[m \; n]'$ can only lie in $S_{01}$ or $S_{11}$.

**Proposition 3**

$e_x \equiv 0$, $\ddot{x}_d \equiv 1$

**Proof**

Since $e_x(0) = 0, \dot{e}_x(0) = 0$, Proposition 3 is proved by Lemma 3 and Lemma 4.

**Proposition 4**

$\Lambda(t) = -\dfrac{\pi}{3}\begin{cases} \text{when } K_{Y_1} \cdot \dot{e}_y + K_{Y_2} \cdot e_y \geqslant -\dfrac{1}{\sqrt{3}}, S_{pq} = S_{10} \\ \text{when } K_{Y_1} \cdot \dot{e}_y + K_{Y_2} \cdot e_y < -\dfrac{1}{\sqrt{3}}, S_{pq} = S_{11} \end{cases}$

(34)

$\Lambda(t) = \dfrac{\pi}{3} : \begin{cases} \text{when } K_{Y_1} \cdot \dot{e}_y + K_{Y_2} \cdot e_y \leqslant \dfrac{1}{\sqrt{3}}, S_{pq} = S_{01} \\ \text{when } K_{Y_1} \cdot \dot{e}_y + K_{Y_2} \cdot e_y > \dfrac{1}{\sqrt{3}}, S_{pq} = S_{11} \end{cases}$

(35)

**Proof**

1° Find $S_{pq}$ when $\Lambda(t) = -\dfrac{\pi}{3}$

To receive $S_{10}$, $[\ddot{x}_d \; \ddot{y}_d]'$ must lie on the red dash line while within the yellow zone or blue zone. In other words, $[\ddot{x}_d \; \ddot{y}_d]'$ should be above point $A$ while on the red dash line in Figure 14.

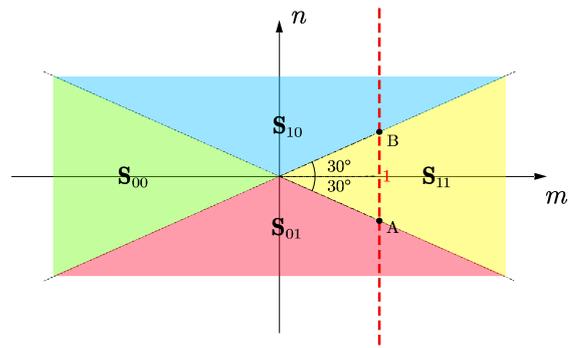

Fig. 14. The possible positions of $[\ddot{x}_d \; \ddot{y}_d]'$.

This requirement can be addressed in (36).
$$\ddot{y}_d \geqslant -\dfrac{1}{\sqrt{3}} \tag{36}$$

Notice that $\ddot{y}_r = 0$. Substituting Equation (26) into Equation (36) yields Equation (37).
$$K_{Y_1} \cdot \dot{e}_y + K_{Y_2} \cdot e_y \geqslant -\dfrac{1}{\sqrt{3}} \tag{37}$$

Similarly, $[\ddot{x}_d \ \ddot{y}_d]'$ should be below point $A$ while on the red dash line in Figure 14 to receive $S_{11}$. It yields Equation (38).

$$K_{Y_1} \cdot \dot{e}_y + K_{Y_2} \cdot e_y < -\frac{1}{\sqrt{3}} \qquad (38)$$

2° Find $S_{pq}$ when $\Lambda(t) = \frac{\pi}{3}$

Similar to the proof in 1°, the desired position of $[\ddot{x}_d \ \ddot{y}_d]'$ now is above or below B in Figure 14. The detail is omitted.

**Remark 3**

Proposition 4 is visualized in Figure 15.

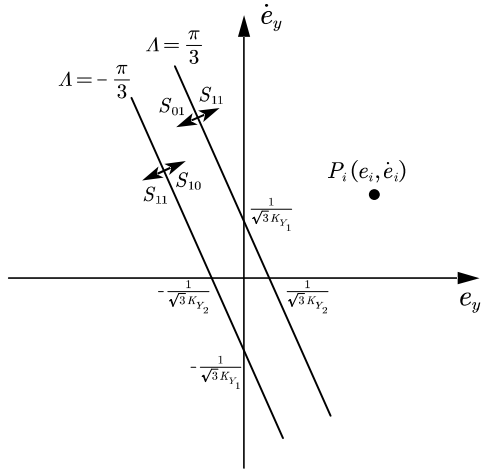

Fig. 15. Proposition 4.

**Lemma 5**

If the system starts from the initial state point $P(e_y, \dot{e}_y) = P_i(e_i, \dot{e}_i)$ satisfying the conditions in (39)

$$\begin{cases} K_{Y_1} \cdot \dot{e}_i + K_{Y_2} \cdot e_i \geq \dfrac{1}{\sqrt{3}} \\ \dot{e}_i \geq 0, \ e_i \geq 0 \\ \Lambda = \dfrac{\pi}{3} \end{cases} \qquad (39)$$

then the state $P(e_y, \dot{e}_y)$ will reach the bound specified in Equation (40) at $t_1$.

$$K_{Y_1} \cdot \dot{e}_y + K_{Y_2} \cdot e_y = \frac{1}{\sqrt{3}} \qquad (40)$$

Further, $0 < t_1 < 1$.

**Proof**

Based on Equation (34), the dynamic state error satisfies Equation (41).

$$\ddot{e}_y = -K_{Y_1} \cdot \dot{e}_y - K_{Y_2} \cdot e_y \qquad (41)$$

Considering the initial condition $P_i(e_i, \dot{e}_i)$, the solution to Equation (41) is (42).

$$\begin{cases} e_y = \left(2e_i + \dfrac{1}{3}\dot{e}_i\right) \cdot e^{-3t} + \left(-e_i - \dfrac{1}{3}\dot{e}_i\right) \cdot e^{-6t} \\ \dot{e}_y = (-6e_i - \dot{e}_i) \cdot e^{-3t} + (6e_i + 2\dot{e}_i) \cdot e^{-6t} \end{cases} \qquad (42)$$

Substituting Equation (42) into Equation (40) yields Equation (43).

$$\frac{1}{9\sqrt{3}} \cdot e^{6t} + \left(2e_i + \frac{1}{3}\dot{e}_i\right) \cdot e^{3t} - 4e_i - \frac{4}{3}\dot{e}_i = 0 \qquad (43)$$

Thus, we have only one positive solution to $e^{3t_1}$. The result is in Equation (44).

$$e^{3t_1} = \frac{-2e_i - \frac{1}{3}\dot{e}_i + \sqrt{\left(2e_i + \frac{1}{3}\dot{e}_i\right)^2 + \frac{4}{9\sqrt{3}} \cdot \left(4e_i + \frac{4}{3}\dot{e}_i\right)}}{\frac{2}{9\sqrt{3}}}$$

$$(44)$$

It can be checked that $0 < t_1 < 1$ if $\dot{e}_i \geq 0$, $e_i \geq 0$.

**Remark 4**

Since the half of the period ($\frac{T}{2}$) lasts 1 second, there is the extra time $t_2$ ($t_2 = \frac{T}{2} - t_1$) after reaching the bound in Equation (40). The dynamic state error will obey the dynamics where $S_{pq} = S_{01}$ for $t_2$. It will not escape from $S_{pq} = S_{01}$ during $t_2$.

**Lemma 6**

If the system starts from the initial state point $P(e_y, \dot{e}_y) = P_i(e_i, \dot{e}_i)$ satisfying conditions in (45)

$$\begin{cases} K_{Y_1} \cdot \dot{e}_i + K_{Y_2} \cdot e_i \leq -\dfrac{1}{\sqrt{3}} \\ \dot{e}_i \leq 0, \ e_i \leq 0 \\ \Lambda = -\dfrac{\pi}{3} \end{cases} \qquad (45)$$

then the state $P(e_y, \dot{e}_y)$ will reach the bound specified in Equation (46) at $t_3$.

$$K_{Y_1} \cdot \dot{e}_y + K_{Y_2} \cdot e_y = -\frac{1}{\sqrt{3}} \qquad (46)$$

Further, $0 < t_3 < 1$.

**Proof**

Similar to the Proof of Lemma 5, $e^{3t_3}$ is calculated in Equation (47).

$$e^{3t_3} = \frac{2e_i + \frac{1}{3}\dot{e}_i + \sqrt{\left(2e_i + \frac{1}{3}\dot{e}_i\right)^2 - \frac{4}{9\sqrt{3}} \cdot \left(4e_i + \frac{4}{3}\dot{e}_i\right)}}{\frac{2}{9\sqrt{3}}}$$

(47)

It can be checked that $0 < t_3 < 1$ if $\dot{e}_i \leq 0$, $e_i \leq 0$.

**Remark 5**

Since the half of the period ($\frac{T}{2}$) lasts 1 second, there is the extra time $t_4$ ($t_4 = \frac{T}{2} - t_3$) after reaching the bound in Equation (46). The dynamic state error will obey the dynamics where $S_{pq} = S_{10}$ for $t_4$. It will not escape from $S_{pq} = S_{10}$ during $t_4$.

**Lemma 7**

If the system starts from initial state $P(e_y, \dot{e}_y) = P_i(e_i, \dot{e}_i)$ satisfying conditions in (39), then the state $P(e_y, \dot{e}_y)$ will arrive at a state satisfying the conditions in (45) after half period (1 second).

**Proof**

We calculate the state after half period (1 second), $P(e_y, \dot{e}_y)|_{t=\frac{T}{2}}$, starting from the initial state $P_i(e_i, \dot{e}_i)$.

During $t_1$ in Equation (44), the dynamic state error obeys equation (30) ($S_{11}$). At time $t_1$, the state reaches the bound defined in Equation (40). After that, the dynamic state error follows Equation (32) ($S_{01}$), which lasts $t_2$ ($t_2 = 1 - t_1$).

The result is plotted in Figure 16. $z-axis$ represents the result of whether $P(e_y, \dot{e}_y)|_{t=\frac{T}{2}}$ satisfies the conditions in (45) in Quadrant III. It can be seen that all the initial states, $P_i(e_i, \dot{e}_i)$, satisfying the conditions in (39), meet this requirement.

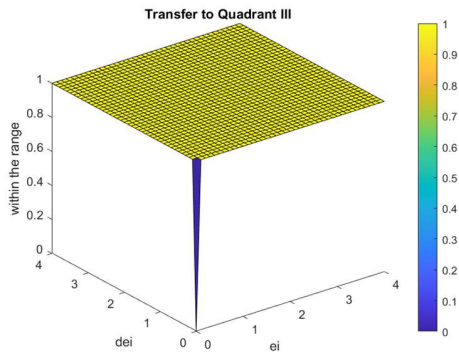

Fig. 16. Proof of Lemma 7.

**Lemma 8**

If the system starts from the initial state point $P(e_y, \dot{e}_y) = P_i(e_i, \dot{e}_i)$ satisfying conditions in (45), then the state $P(e_y, \dot{e}_y)$ will arrive at a state satisfying conditions in (39) after half period (1 second).

**Proof**

Similar to the Proof of Lemma 7, we find that all the initial states satisfying the condition in (45) will arrive at a state satisfying conditions in (39) after half period. It can be checked in Figure 17.

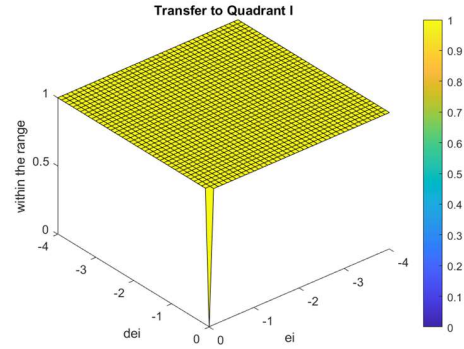

Fig. 17. Proof of Lemma 8.

**Proposition 5**

If an initial state, $P_i(e_i, \dot{e}_i)$, satisfies the conditions in (48)

$$\begin{cases} K_{Y_1} \cdot \dot{e}_i + K_{Y_2} \cdot e_i \leq -\frac{1}{\sqrt{3}} \\ \dot{e}_i \leq 0,\ e_i \leq 0 \\ \Lambda = -\frac{\pi}{3} \end{cases} \quad or \quad \begin{cases} K_{Y_1} \cdot \dot{e}_i + K_{Y_2} \cdot e_i \geq \frac{1}{\sqrt{3}} \\ \dot{e}_i \geq 0,\ e_i \geq 0 \\ \Lambda = \frac{\pi}{3} \end{cases}$$

(48)

then $P(e_y, \dot{e}_y)|_{t=n \cdot \frac{T}{2}}$ $(n \in \mathbf{Z}^+)$ also satisfies the conditions in (48) applying the controller we set.

**Proof**

It can be proved from Lemma 7 and Lemma 8, recursively.

**Remark 6**

Proposition 5 can be explained in Figure 18. Once the initial state is from the yellow area or the blue area, the states after time $t = n \cdot \frac{T}{2}$ $(n \in \mathbf{Z}^+)$ will also be inside yellow area or the blue area.

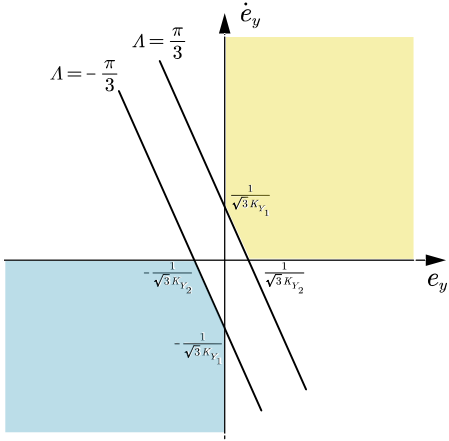

Fig. 18. The area captures the states (Proposition 5).

**Lyapunov Candidate**

Define the Lyapunov candidate in Equation (49)

$$\mathscr{L} = \frac{1}{2} \cdot \dot{e}_y^2 + \frac{1}{2} \cdot K_{Y_2} \cdot e_y^2 \tag{49}$$

Instead of proving $\dot{\mathscr{L}}$ is negative/nonpositive (actually it is not), we will prove that $\mathscr{L}$ is bounded in the rest of this section.

**Property of The Defined Lyapunov Candidate**

We analyze the behavior of Lyapunov candidate for each $S_{pq}$.

$1°\ S_{pq} = S_{11}$

Similar to Equation (17), (18), we receive (50).

$$\dot{\mathscr{L}} = -K_{Y_1} \cdot \dot{e}_y^2 \leqslant 0 \tag{50}$$

$2°\ S_{pq} = S_{10}$

From (31), considering $e_x \equiv 0$, we receive (51).

$$\begin{cases} \ddot{e}_y = \dfrac{1}{\sqrt{3}} \\ \dot{e}_y = \dfrac{1}{\sqrt{3}} \cdot t + \dot{e}_{y_0} \\ e_y = \dfrac{1}{2\sqrt{3}} \cdot t^2 + \dot{e}_{y_0} \cdot t + e_{y_0} \end{cases} \tag{51}$$

where $e_{y_0}$ and $\dot{e}_{y_0}$ are the initial dynamic state errors (constant).

Thus, the relevant Lyapunov candidate is in (52).

$$\mathscr{L} = \frac{1}{2} \cdot \left(\frac{1}{\sqrt{3}} \cdot t + \dot{e}_{y_0}\right)^2 \\ + \frac{1}{2} \cdot K_{Y_2} \cdot \left(\frac{1}{2\sqrt{3}} \cdot t^2 + \dot{e}_{y_0} \cdot t + e_{y_0}\right)^2 \tag{52}$$

$3°\ S_{pq} = S_{01}$

From (32), considering $e_x \equiv 0$, we receive (53).

$$\begin{cases} \ddot{e}_y = -\dfrac{1}{\sqrt{3}} \\ \dot{e}_y = -\dfrac{1}{\sqrt{3}} \cdot t + \dot{e}_{y_0} \\ e_y = -\dfrac{1}{2\sqrt{3}} \cdot t^2 + \dot{e}_{y_0} \cdot t + e_{y_0} \end{cases} \tag{53}$$

where $e_{y_0}$ and $\dot{e}_{y_0}$ are the initial dynamic state errors (constant).

Thus, the relevant Lyapunov candidate is in (54).

$$\mathscr{L} = \frac{1}{2} \cdot \left(-\frac{1}{\sqrt{3}} \cdot t + \dot{e}_{y_0}\right)^2 \\ + \frac{1}{2} \cdot K_{Y_2} \cdot \left(-\frac{1}{2\sqrt{3}} \cdot t^2 + \dot{e}_{y_0} \cdot t + e_{y_0}\right)^2 \tag{54}$$

**Proposition 6**

If an initial state, $P_i(e_i, \dot{e}_i)$, satisfies the conditions defined in (55),

$$\begin{cases} K_{Y_1} \cdot \dot{e}_i + K_{Y_2} \cdot e_i \leqslant -\dfrac{1}{\sqrt{3}} \\ \dot{e}_i \leqslant 0,\ e_i \leqslant 0 \\ \varLambda = -\dfrac{\pi}{3} \end{cases} or \begin{cases} K_{Y_1} \cdot \dot{e}_i + K_{Y_2} \cdot e_i \geqslant \dfrac{1}{\sqrt{3}} \\ \dot{e}_i \geqslant 0,\ e_i \geqslant 0 \\ \varLambda = \dfrac{\pi}{3} \end{cases} \tag{55}$$

applying the controller we set, we have the result in (56).

$$\mathscr{L}(t) \leqslant \max\left\{\mathscr{L}\left(n \cdot \frac{T}{2}\right), \mathscr{L}\left((n+1) \cdot \frac{T}{2}\right)\right\}, \\ n \in \mathbf{Z}^+,\ t \in \left[n \cdot \frac{T}{2},\ (n+1) \cdot \frac{T}{2}\right] \tag{56}$$

**Proof**

$1°$ The initial state, $P_i(e_i, \dot{e}_i)$, satisfies the first condition set in (55)

The state $P(e_y, \dot{e}_y)$ is initially inside the area dominated by the switch matrix $S_{11}$. This phase lasts $t_3$ in (47). The change of the Lyapunov candidate obeys (50).

Thus, we have the result in (57).

$$\mathscr{L}(t) = \mathscr{L}\left(n \cdot \frac{T}{2}\right) + \int_{n \cdot \frac{T}{2}}^{t} \dot{\mathscr{L}}(t) \leqslant \mathscr{L}\left(n \cdot \frac{T}{2}\right), \\ n \in \mathbf{Z}^+,\ t \in \left[n \cdot \frac{T}{2},\ n \cdot \frac{T}{2} + t_3\right] \tag{57}$$

We can also find the minimum Lyapunov candidate during $t \in \left[n \cdot \dfrac{T}{2},\ n \cdot \dfrac{T}{2} + t_3\right]$ in (58).

$$\min\{\mathscr{L}(t)\} = \mathscr{L}\left(n \cdot \frac{T}{2} + t_3\right),$$
$$n \in \mathbf{Z}^+, \ t \in \left[n \cdot \frac{T}{2}, \ n \cdot \frac{T}{2} + t_3\right] \tag{58}$$

At time $n \cdot \frac{T}{2} + t_3$, the state $P(e_y, \dot{e}_y)$ reaches the bound and enters the area dominated by the switch matrix $S_{10}$. For the rest of the time ($t_4 = \frac{T}{2} - t_3$), the Lyapunov candidate follows the rule in (52).

Denote the state $P(e_y, \dot{e}_y)$ at $n \cdot \frac{T}{2} + t_3$ in (59).

$$\left. e_y \right|_{t = n \cdot \frac{T}{2} + t_3} = e_M$$
$$\left. \dot{e}_y \right|_{t = n \cdot \frac{T}{2} + t_3} = \dot{e}_M \tag{59}$$

Substituting (59) into (46) yields (60).

$$\ddot{e}_M + 2 e_M = -\frac{1}{9\sqrt{3}} \tag{60}$$

Substituting (59), (60) into (54) and calculating its derivative yields (61).

$$\dot{\mathscr{L}} = 9 \cdot \left(\frac{1}{\sqrt{3}} \cdot t + \dot{e}_M\right) \cdot \left(\frac{1}{\sqrt{3}} \cdot t^2 + 2 \cdot \dot{e}_M \cdot t - \dot{e}_M\right) \tag{61}$$

Based on the derivative in Equation (61), we receive the monotone of $\mathscr{L}$ below.

a. when $\dot{e}_M \in \left(-\infty, -\frac{1}{\sqrt{3}}\right) \cup \left(-\frac{1}{\sqrt{3}}, 0\right) \cup (0, +\infty)$

During $t_4$, $\mathscr{L}$ decreases monotonically / $\mathscr{L}$ decreases for a while and increases later.

b. when $\dot{e}_M = 0$

During $t_4$, $\mathscr{L}$ increases monotonically.

c. when $\dot{e}_M = -\frac{1}{\sqrt{3}}$

During $t_4$, $\mathscr{L}$ decreases monotonically.

Thus, we make a conclusion in (62).

$$\mathscr{L}(t) \leqslant \max\left\{\mathscr{L}\left(n \cdot \frac{T}{2} + t_3\right), \ \mathscr{L}\left((n+1) \cdot \frac{T}{2}\right)\right\},$$
$$n \in \mathbf{Z}^+, \ t \in \left[n \cdot \frac{T}{2} + t_3, \ (n+1) \cdot \frac{T}{2}\right] \tag{62}$$

Substituting the relationship in (58) into (62), we receive the result in (56).

2° The initial state, $P_i(e_i, \dot{e}_i)$, satisfies the second condition set in (55)

The state $P(e_y, \dot{e}_y)$ is initially inside the area dominated by the switch matrix $S_{11}$. This phase lasts $t_1$ in (44). The change of the Lyapunov candidate obeys (50).

Thus, we make a conclusion in (63).

$$\mathscr{L}(t) = \mathscr{L}\left(n \cdot \frac{T}{2}\right) + \int_{n \cdot \frac{T}{2}}^{t} \dot{\mathscr{L}}(t) \leqslant \mathscr{L}\left(n \cdot \frac{T}{2}\right),$$
$$n \in \mathbf{Z}^+, \ t \in \left[n \cdot \frac{T}{2}, \ n \cdot \frac{T}{2} + t_1\right] \tag{63}$$

We also find the minimum Lyapunov candidate during $t \in \left[n \cdot \frac{T}{2}, \ n \cdot \frac{T}{2} + t_1\right]$ in (64).

$$\min\{\mathscr{L}(t)\} = \mathscr{L}\left(n \cdot \frac{T}{2} + t_1\right),$$
$$n \in \mathbf{Z}^+, \ t \in \left[n \cdot \frac{T}{2}, \ n \cdot \frac{T}{2} + t_1\right] \tag{64}$$

At time $n \cdot \frac{T}{2} + t_1$, the state $P(e_y, \dot{e}_y)$ hits the bound and enters the area dominated by the switch matrix $S_{01}$. For the rest of the time ($t_2 = \frac{T}{2} - t_1$), the Lyapunov candidate follows the rule in (54).

Denote the state $P(e_y, \dot{e}_y)$ at $n \cdot \frac{T}{2} + t_1$ in (65).

$$\left. e_y \right|_{t = n \cdot \frac{T}{2} + t_1} = e_m$$
$$\left. \dot{e}_y \right|_{t = n \cdot \frac{T}{2} + t_1} = \dot{e}_m \tag{65}$$

Substituting (65) into (40) yields (66).

$$\ddot{e}_m + 2 e_m = \frac{1}{9\sqrt{3}} \tag{66}$$

Substituting (65), (66) into (52) and calculating its derivative yields (67).

$$\dot{\mathscr{L}} = 9 \cdot \left(\frac{1}{\sqrt{3}} \cdot t - \dot{e}_m\right) \cdot \left(\frac{1}{\sqrt{3}} \cdot t^2 - 2 \cdot \dot{e}_m \cdot t + \dot{e}_m\right) \tag{67}$$

Based on the derivative in (67), we receive the monotone of $\mathscr{L}$ below.

a. when $\dot{e}_m \in (-\infty, 0) \cup \left(0, \frac{1}{\sqrt{3}}\right) \cup \left(\frac{1}{\sqrt{3}}, +\infty\right)$

During $t_2$, $\mathscr{L}$ decreases monotonically / $\mathscr{L}$ decreases for a while and increases later.

b. when $\dot{e}_m = 0$

During $t_2$, $\mathscr{L}$ increases monotonically.

c. when $\dot{e}_m = \frac{1}{\sqrt{3}}$

During $t_2$, $\mathscr{L}$ decreases monotonically.

Thus, we find the following relationship in (68).

$$\mathcal{L}(t) \leqslant \max\left\{\mathcal{L}\left(n \cdot \frac{T}{2} + t_1\right), \mathcal{L}\left((n+1) \cdot \frac{T}{2}\right)\right\}, \quad (68)$$
$$n \in \mathbf{Z}^+, \ t \in \left[n \cdot \frac{T}{2} + t_1, (n+1) \cdot \frac{T}{2}\right]$$

Substituting (64) into (68), we receive (56).

**Remark 7**

Proposition 6 tells us that the Lyapunov candidate is locally maximized at $t = n \cdot \frac{T}{2}$ ($n \in \mathbf{Z}^+$) if meeting the requirements in (55).

It can be demonstrated in Figure 19. When $t \geqslant \frac{T}{2}$, $\mathcal{L}(t)$ is bounded by $\mathcal{L}\left(n \cdot \frac{T}{2}\right)$, $(n \in \mathbf{Z}^+)$.

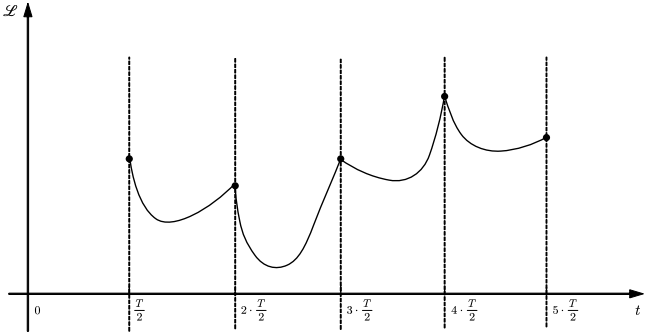

Fig. 19. Lyapunov candidate at $t = n \cdot \frac{T}{2}$, $(n \in \mathbf{Z}^+)$, is locally maximized

**Remark 8 (First Half Period)**

The behavior of the state $P(e_y, \dot{e}_y)$ during $\left[0, \frac{T}{2}\right]$ follows (51), the rule for $S_{10}$, with the initial conditions $e_y|_{t=0} = 0$ and $\dot{e}_y|_{t=0} = 0$. The Lyapunov candidate in bounded during the first half period. The state $P(e_y, \dot{e}_y)$ is within space defined by (39) at $\frac{T}{2}$.

**Inference 1 (Stability Criteria)**

If $\mathcal{L}\left(n \cdot \frac{T}{2}\right)$, $(n \in \mathbf{Z}^+)$, is upper bounded and $\mathcal{L}(t)$, $\left(t \in \left[0, \frac{T}{2}\right]\right)$, is upper bounded,
then $\mathcal{L}(t)$ is bounded.

**Inference 2 (Lyapunov Candidate Upper Bound)**

If $\mathcal{L}(t)$ is upper bounded,
then we can find the upper bound in (69).

$$\mathcal{L}(t) \leqslant \max_{n \in \mathbf{Z}^+, n \geqslant C}\left\{\mathcal{L}\left(n \cdot \frac{T}{2}\right)\right\}, \quad (69)$$
$$\left(t \geqslant C \cdot \frac{T}{2}, \ C \text{ is a positive constant}\right)$$

**Lemma 9**

Define $_\Delta \mathcal{L}(n)$ in (70).

$$_\Delta \mathcal{L}(n) = \mathcal{L}\left((n+1) \cdot \frac{T}{2}\right) - \mathcal{L}\left(n \cdot \frac{T}{2}\right), \quad (70)$$
$$(n \in \mathbf{Z}^+)$$

$_\Delta \mathcal{L}(n)$ is determined by $e_y|_{t=n \cdot \frac{T}{2}}$ and $\dot{e}_y|_{t=n \cdot \frac{T}{2}}$. The relationship between $(e_y, \dot{e}_y)|_{t=n \cdot \frac{T}{2}}$ and $_\Delta \mathcal{L}(n)$ is plotted in Figure 20, 21.

The cases where $(e_y, \dot{e}_y)|_{t=n \cdot \frac{T}{2}}$ is in Quadrant I and Quadrant III are demonstrated in Figure 20 and Figure 21, respectively.

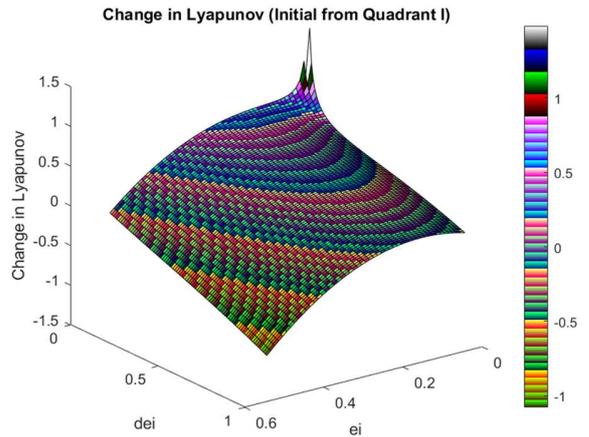

Fig. 20. $(e_y, \dot{e}_y)|_{t=n \cdot \frac{T}{2}}$ in Quadrant I.

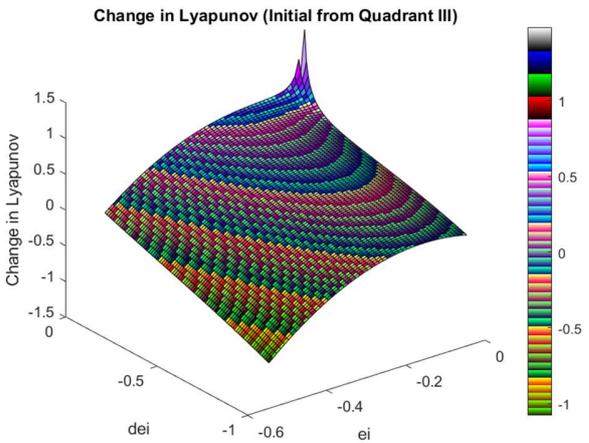

Fig. 21. $(e_y, \dot{e}_y)|_{t=n \cdot \frac{T}{2}}$ in Quadrant III.

**Remark 9**

Parts of the area in Figure 20 and 21 are unnecessary. The area we are interested in are the colored zones in Figure 18, defined in (48).

One might be depressed from finding positive $_\Delta\mathscr{L}(n)$ in some initial conditions $(e_y, \dot{e}_y)|_{t=n\cdot\frac{T}{2}}$ from Figure 20 and Figure 21. These initial conditions indicate the failure of proving asymptotic stable by this method. However, this system is not asymptotically stable.

**Remark 10 (Upper Bound of $_\Delta\mathscr{L}(n)$)**

It can be concluded from Figure 20 and Figure 21 that the maximum value of $_\Delta\mathscr{L}(n)$ is either on the bound defined in (40) or on the bound defined in (46)

Calculating the maximum of $_\Delta\mathscr{L}(n)$, we receive the same result on either bound. That is (71).

$$_\Delta\mathscr{L}(n) \leqslant \frac{3}{4} \tag{71}$$

**Remark 11 (Graph of Positive $_\Delta\mathscr{L}(n)$)**

We can also plot figure 20 and Figure 21 in 2 dimensions. The 2-D graph are in Figure 22 and 23.

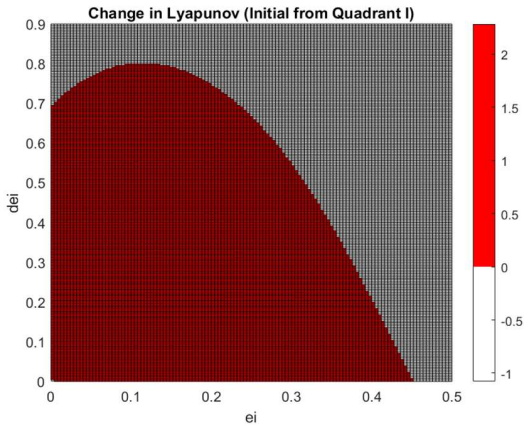

Fig. 22. $(e_y, \dot{e}_y)|_{t=n\cdot\frac{T}{2}}$ in Quadrant I.

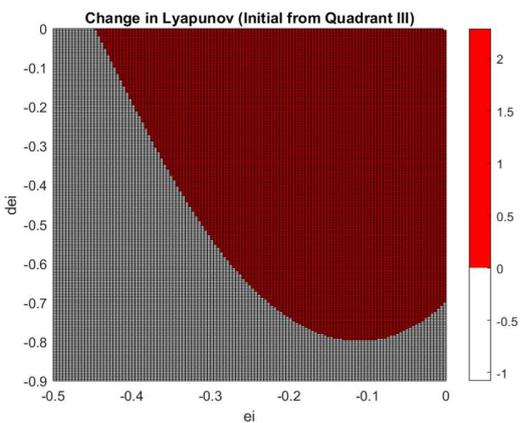

Fig. 23. $(e_y, \dot{e}_y)|_{t=n\cdot\frac{T}{2}}$ in Quadrant III.

The red area in Figure 22 and 23 is the $(e_y, \dot{e}_y)|_{t=n\cdot\frac{T}{2}}$ receiving a positive $_\Delta\mathscr{L}(n)$, increasing $\mathscr{L}\left(n\cdot\frac{T}{2}\right), (n\in\mathbf{Z}^+)$. The white area, on the other hand, receives a negative $_\Delta\mathscr{L}(n)$, decreasing $\mathscr{L}\left(n\cdot\frac{T}{2}\right), (n\in\mathbf{Z}^+)$.

**Main Proof**

Since the first half period (See Remark 8) is bounded and lets the $(e_y, \dot{e}_y)|_{t=\frac{T}{2}}$ be inside the zone defined by (39), $(e_y, \dot{e}_y)|_{t=n\cdot\frac{T}{2}}$ will always be inside (See Remark 6) the zone defined by (55), which is the colored area in Figure 18.

The $\mathscr{L}\left(n\cdot\frac{T}{2}\right), (n\in\mathbf{Z}^+)$ defined is bounded. The reason is given below:

$1°$ When $(e_y, \dot{e}_y)|_{t=i\cdot\frac{T}{2}}, (i\in\mathbf{Z}^+)$ drops inside the white area in Figure 22 or Figure 23.

$_\Delta\mathscr{L}(i) < 0$ (See Remark 11). Thus, we have (72).

$$(e_y, \dot{e}_y)|_{t=i\cdot\frac{T}{2}} < (e_y, \dot{e}_y)|_{t=(i+1)\cdot\frac{T}{2}} \tag{72}$$

$2°$ When $(e_y, \dot{e}_y)|_{t=i\cdot\frac{T}{2}}, (i\in\mathbf{Z}^+)$ drops inside the red area in Figure 22 or Figure 23.

$_\Delta\mathscr{L}(i) > 0$ (See Remark 11). It tries to push $(e_y, \dot{e}_y)|_{t=(i+1)\cdot\frac{T}{2}}$ outside the red area.

Notice that $_\Delta\mathscr{L}(i)$ has an upper bound (See Remark 10). It ensures that $(e_y, \dot{e}_y)|_{t=(i+1)\cdot\frac{T}{2}}$ is bounded for this case.

$3°$ When $(e_y, \dot{e}_y)|_{t=i\cdot\frac{T}{2}}, (i\in\mathbf{Z}^+)$ drops inside the rest area where $_\Delta\mathscr{L}(i) = 0$.

$(e_y, \dot{e}_y)|_{t=(i+1)\cdot\frac{T}{2}} = (e_y, \dot{e}_y)|_{t=i\cdot\frac{T}{2}}$.

Since $(e_y, \dot{e}_y)|_{t=\frac{T}{2}}$ is bounded, $\mathscr{L}\left(n\cdot\frac{T}{2}\right), (n\in\mathbf{Z}^+)$ is also bounded, proved recursively by $1°$, $2°$, and $3°$ above.

From Inference 1, we can make the conclusion — $\mathscr{L}(t)$ is bounded.

The proof is complete.

## 7. Simulation Results

The dynamic state error result is in Figure 24. The dynamic state error along $x-axis$ remains zero. The dynamic state error along $y-axis$ is not stabilized at zero. Although it is not asymptotically stable, the dynamic state error along $y-axis$ is bounded.

Figure 25 displays the history of inputs. It can be seen that the inputs change periodically in general. The period is identical to the period ($T$) of the gait planned. Saturation happens from time to time. While there are the time windows where no saturation appears.

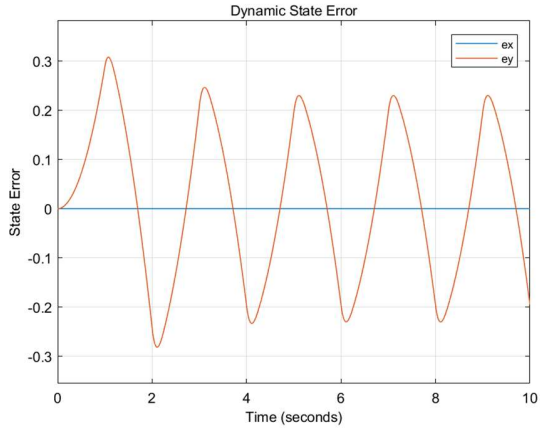

Fig. 24. Dynamic state errors in 2 dimensions

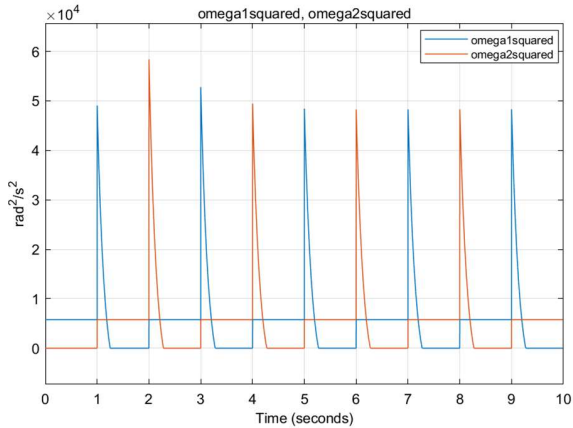

Fig. 25. Inputs

Define the angle of acceleration $\left(\ddot{\ddot{u}}, \ddot{\ddot{v}}\right)$ in (73). Notice the angle is converted into the range $[0, 2\pi]$.

$$angle\ of\ \left(\ddot{\ddot{u}}, \ddot{\ddot{v}}\right) = 0\ to\ 2\pi\ \{\operatorname{atan2}\left(\ddot{\ddot{v}}, \ddot{\ddot{u}}\right)\} \quad (73)$$

We can also check whether the saturation happens in Figure 26. The inputs for the tilt vehicle avoid the saturation if and only if the yellow curve, which represents the desired angle of acceleration, is between the purple curve, which represents the upper bound of the possible angle of acceleration, and the green curve, which represents the lower bound of the possible angle of acceleration.

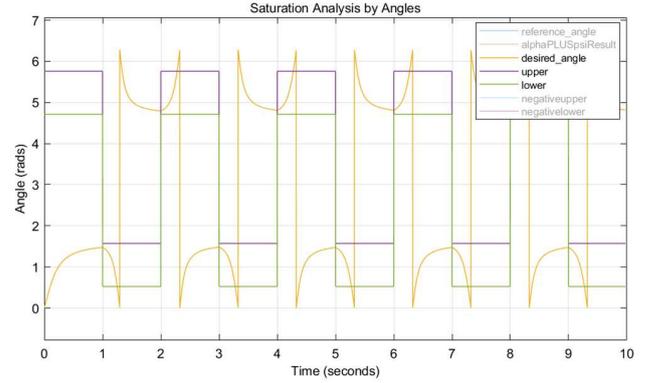

Fig. 26. The angle of acceleration.

The history of the Lyapunov candidate is diagramed in Figure 27. It is not negative-definite, failing to prove the asymptotically stable. While the system is not asymptotically stable judged from Figure 24.

On the other hand, the Lyapunov candidate is bounded in Figure 27. The value of Lyapunov candidate is locally maximized at time $t = n \cdot \dfrac{T}{2}$ , $(n \in \mathbf{Z}^+)$ . It is as expected in Remark 7.

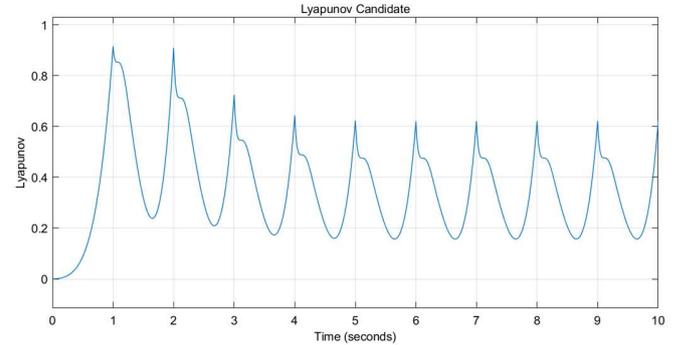

Fig. 27. The record of the Lyapunov candidate.

## 8. Conclusions and Discussions

Two different penguin-inspired gaits are applied to facilitate to control the tilt vehicle in this research. The first gait introduces no saturation in control input, reaching the zero dynamic state error in the end. The controller in the second gait fails to asymptotically stabilize the tilt vehicle. Saturation happens in this case. While the state error in the second gait case is bounded. The stability proofs for both cases are revealed. We picked the same Lyapunov candidate for both cases. It is proved semi-negative-definite for the first gait. While it is proved bounded for the second case.

In 'Main Proof' in Section 6, we proved that $\mathscr{L}(t)$ is bounded. It can be interesting to address further discussions on the upper bound of the Lyapunov candidate in the second case (gait causing saturation).

The Lyapunov candidate in (49) can be rewritten in (74), substituting $K_{Y_2}$.

$$\mathscr{L} = \frac{1}{2} \cdot \dot{e}_y{}^2 + 9 \cdot e_y{}^2 \qquad (74)$$

(74) can be further written in (75).

$$\frac{e_y{}^2}{\left(\frac{\sqrt{\mathscr{L}}}{3}\right)^2} + \frac{\dot{e}_y{}^2}{\left(\sqrt{2} \cdot \sqrt{\mathscr{L}}\right)^2} = 1 \qquad (75)$$

(75) defines an ellipse in Figure 28.

When we receive a particular $\mathscr{L}$, the corresponding $(e_y, \dot{e}_y)$ must fall on the edge of this ellipse. It is also worth noting that $(e_y, \dot{e}_y)|_{t = n \cdot \frac{T}{2}}$, $(n \in \mathbf{Z}^+)$, can only appear on the edge of the ellipse in Quadrant I and Quadrant III (See Remark 6).

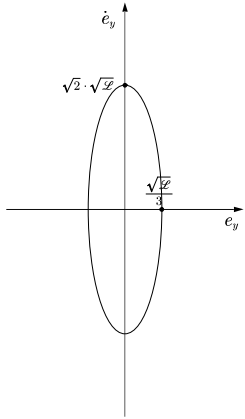 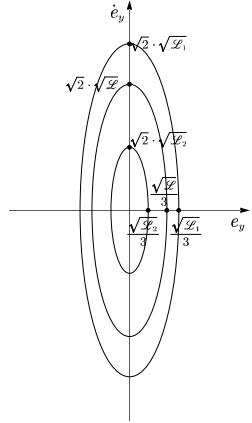

Fig. 28. Possible $(e_y, \dot{e}_y)$.   Fig. 29. different $\mathscr{L}$

When we receive a larger value in Lyapunov candidate (e.g., $\mathscr{L}_1 > \mathscr{L}$), the ellipse is also expanded (See the $\mathscr{L}_1 - induced$ ellipse in Figure 29). When we receive a smaller Lyapunov (e.g., $\mathscr{L}_2 < \mathscr{L}$), the ellipse is also shrunk (See the $\mathscr{L}_2 - induced$ ellipse in Figure 29).

Define $\mathscr{L}_{critical}$ in (76).

$$\begin{cases} \{\mathscr{L}_{critical} - induced\ ellipse\} \cap \left\{(e_y, \dot{e}_y)\right\}_{\triangle \mathscr{L}(n) \geqslant 0} \neq \varnothing \\ \\ \forall \mathscr{L} > \mathscr{L}_{critical}: \\ \{\mathscr{L} - induced\ ellipse\} \cap \left\{(e_y, \dot{e}_y)\right\}_{\triangle \mathscr{L}(n) \geqslant 0} = \varnothing \end{cases} \qquad (76)$$

where $\left\{(e_y, \dot{e}_y)\right\}_{\triangle \mathscr{L}(n) \geqslant 0}$ is the collection of $(e_y, \dot{e}_y)$ satisfying $\triangle \mathscr{L}(n) \geqslant 0$ (See Lemma 9).

Thus, $\left\{(e_y, \dot{e}_y)\right\}_{\triangle \mathscr{L}(n) \geqslant 0}$ is (part of) the red area in Figure 22 and Figure 23 (See Remark 11).

Figure 30 plots the $\mathscr{L}_{critical} - induced$ ellipse.

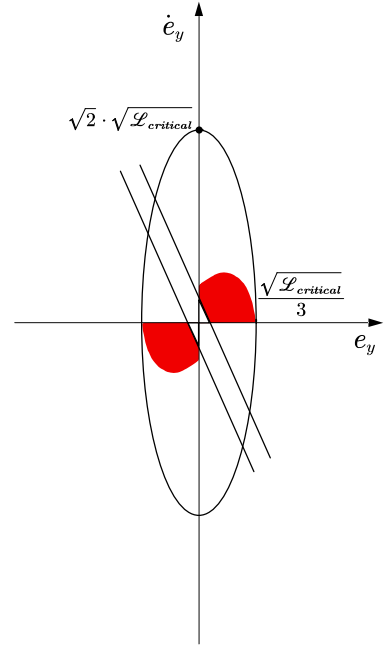

Fig. 30. $\mathscr{L}_{critical} - induced$ ellipse.

**Proposition 7**

$$supremum(\mathscr{L}(t)) \leqslant \mathscr{L}_{critical} + \frac{3}{4},$$
$$(t \geqslant t_L,\ t_L\ is\ a\ finite\ const) \qquad (77)$$

**Proof**

We set $t_L$ sufficiently large so that $(e_y, \dot{e}_y)|_{t = \rho \cdot \frac{T}{2}}$, $\left(\rho \in \mathbf{Z}^+, \rho \cdot \frac{T}{2} < t_L\right)$, falls inside the red area defined in Figure 22 and Figure 23.

1° $\forall n > \rho$, $(n \in \mathbf{Z}^+)$, $(e_y, \dot{e}_y)|_{t = n \cdot \frac{T}{2}}$ is unable to enter the white area in Figure 22 and Figure 23

In this case, we have (78).

$$supremum(\mathscr{L}(t)) \leqslant \mathscr{L}_{critical} < \mathscr{L}_{critical} + \frac{3}{4},$$
$$(t \geqslant t_L) \qquad (78)$$

2° $\exists m > \rho$, $(m \in \mathbf{Z}^+)$, $(e_y, \dot{e}_y)|_{t = m \cdot \frac{T}{2}}$ enters the white area while $(e_y, \dot{e}_y)|_{t = (m-1) \cdot \frac{T}{2}}$ is in the red area

In this case, we have (79).

$$\mathscr{L}\left(m \cdot \frac{T}{2}\right) = \mathscr{L}\left((m-1) \cdot \frac{T}{2}\right) + {}_\Delta\mathscr{L}(m-1)$$
$$\leqslant \mathscr{L}_{critical} + \frac{3}{4} \quad (79)$$

After entering the white area, we have ${}_\Delta\mathscr{L} < 0$. We may have case $2°a$ or case $2°b$ subsequently.

$2°a$. If $\forall q > m$, $(q \in \mathbf{Z}^+)$, $(e_y, \dot{e}_y)\big|_{t = q \cdot \frac{T}{2}}$ is unable to enter the red area.

$$\mathscr{L}\left(q \cdot \frac{T}{2}\right) < \mathscr{L}\left(m \cdot \frac{T}{2}\right) \leqslant \mathscr{L}_{critical} + \frac{3}{4} \quad (80)$$

Thus,

$$supremum(\mathscr{L}(t)) \leqslant \mathscr{L}\left(m \cdot \frac{T}{2}\right)$$
$$< \mathscr{L}_{critical} + \frac{3}{4}, \quad (t \geqslant t_L) \quad (81)$$

$2°b$. $\exists w > m$, $(w \in \mathbf{Z}^+)$, $(e_y, \dot{e}_y)\big|_{t = w \cdot \frac{T}{2}}$ enters the red area while $(e_y, \dot{e}_y)\big|_{t = (w-1) \cdot \frac{T}{2}}$ is in the white area $(e_y, \dot{e}_y)\big|_{t = w \cdot \frac{T}{2}}$ returns to the red area where $(e_y, \dot{e}_y)\big|_{t = \rho \cdot \frac{T}{2}}$ is in. This case is recursively proved by $1°$ and $2°a$.

It is worth mentioning that Proposition 7 provides a relaxed upper bound for Lyapunov candidate. Further approaching the supremum is beyond the main scope of this research.

The actual $(e_y, \dot{e}_y)$ history in the result is plotted in Figure 31.

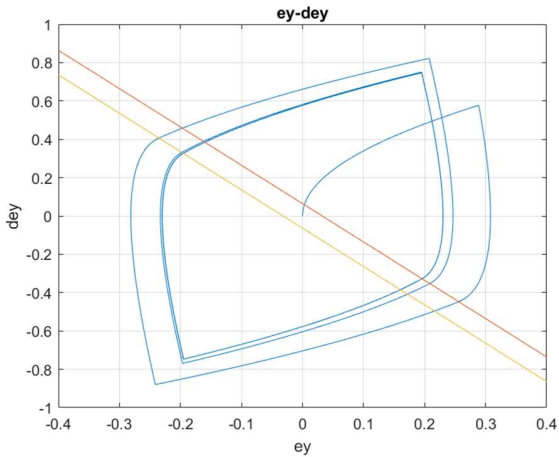

Fig. 31. $(e_y, \dot{e}_y)$ history

Another point worth mentioning is that both the PD control coefficients (coefficients in (9) and (10)) and the planned gait (period, $T$, and the amplitude, $\Lambda$) can affect the stability. For some controller settings, the stability proof forwarded in this research can be challenged (some cases even cause the system unstable. That is partially the reason we avoid deducing the stability proof in a totally analytical way.

One further step can be advocating the controller to a more complicated reference (e.g., circular). Another is to generalize our stability proof to be applicable to wider control parameters.